\documentclass[a4paper,UKenglish]{lipics}
\usepackage{cite}
\usepackage{microtype}

\title{{\sf RAESON: }A Tool for Reasoning Tasks Driven by Interactive Visualization of Logical Structure}
\titlerunning{{\sf RAESON: }A Software Tool for Interactively Visualizing Logical Structure} 
\author{\c Stefan Minic\u a}
\affil{\texttt{stefan.minica@gmail.com}}
\authorrunning{\c Stefan Minic\u a} 

\Copyright{\c Stefan Minic\u a}


\serieslogo{logo_ttl}
\volumeinfo
  {M. Antonia {Huertas}, Jo\~ao {Marcos}, Mar\'ia {Manzano}, Sophie {Pinchinat}, \\
  Fran\c{c}ois {Schwarzentruber}}
  {5}
  {4th International Conference on Tools for Teaching Logic}
  {1}
  {1}
  {147}
\EventShortName{TTL2015}

\begin{document}

\maketitle

\begin{abstract}
The paper presents a software tool for analysis and interactive engagement in various logical reasoning tasks. 
A first feature of the program consists in providing an interface for working with logic-specific repositories of formal knowledge. 
A second feature provides the means to intuitively visualize and interactively generate the 
underlying logical structure that propels customary logical reasoning tasks. 
Starting from this we argue that both aspects have didactic potential and can be integrated in 
teaching activities to provide an engaging learning experience.
\end{abstract}

\section{Introduction}
The tradition of using diagrammatic representations of formal structure in logical reasoning stretches back to 
times when sand or papyrus were the media for visualizations.  
The tradition of using interactive software tools in logical reasoning is a recent refinement of an ancient interest. 
This stretches back to the age of compact disk distributed software alongside paper printed books, 
with~\cite{LPL} as a paradigmatic example, and plenty others, many of which have been presented at this venue~\cite{TTL2}.
Using didactic software online for logic education and e-learning is an equally respectable tradition going back to 
the age when java applets roamed on the web, with~\cite{AproS} as a representative example, and plenty others, 
many of which have been also presented at this venue~\cite{TTL3}. 
Latest trends in this evolution are dissemination of logic courseware freely available online, 
with~\cite{lia} as a pertinent example, alongside didactic software tools built on a technology stack up to date with 
current web standards, with~\cite{MLP} as a relevant illustration. 
This is the general trend for many other fields across the curriculum~\cite{KhanAcad}.
When compared to what exists in other fields, many reasoning and modeling techniques emerging in current 
branches of logic remain in many ways still underrepresented, e.g.  modal logics, interrogative logics, epistemic logics, etc.
However, consult~\cite{mettel,demo,lotrec,tabworkbench,intohylo,leantap} as 
a short list of notable exceptions containing robust software tools for reasoning and modeling 
techniques emerging in nonclassical logics that have been recently integrated into this tradition, 
frequently accompanied by  web workers or didactically tailored online demos. 
Inside this general landscape, the paper will revisit classical challenges for teaching logic and reasoning skills and
implement possibly new software tools for interactive reasoning tasks
driven by automated visualization of logical structure 
emerging in logics that are often underrepresented in or overlooked by existing approaches.  

\textbf{Overview} The paper is structured as follows: 
in the first section we introduce the topic and describe the main architecture of the program and some design options.
Section~\ref{sec:managing-repositories} describes the use of the repositories of formal knowledge underlying the program.
Section~\ref{sec:interactive-visualizations} describes the reasoning tasks and interactive visualization tools provided by the software.
Sections~\ref{sec:dydactic-touch} and~\ref{sec:classroom} explore teaching use case scenarios and provide evidence for didactic relevance.
Section~\ref{sec:conclusion} draws conclusions and hints towards possible directions for further work.

\textbf{The Overall Architecture}
The rest of this section contains information about implementation and development aspects. 
Readers only interested in using the program will probably want to skip to Section \ref{sec:managing-repositories} 
where the program's features are described. 

\textit{The Backend Repositories of Formal Knowledge}
The program consists of two layers: remote repositories of formal knowledge are stored and managed on the server side 
and the client side consists of the user interface to interact with and visualize the repositories' formal content. 
The backend application layer is a REST-like API built using NodeJS 
and Express to provide the usual CRUD operations on repository items. The storage layer consists of several document 
structured databases built using MongoDB 
and Mongoose to define repository items~\cite{NodeJS-Express-MongoDB}.
The databases contain items from the standard domains in a logic curriculum: 
propositional logic, modal logic, predicate logic as well as quantified modal logic. 
Database entries are JSON documents 
with each item containing a basic ASCII representation of its content, 
as received from an input device~\cite{JSON-ascii}, together with a set of additional 
representation formats (MathJax-\LaTeX, D3-SVG, Unicode, etc.) used in visualizations, 
interactive reasoning tasks and natural language information.
The left and middle panels in Figure~\ref{figure:main+editor} illustrate intuitively the list representation and 
the item features for the formulae library database. Section~\ref{sec:managing-repositories} contains further information 
about how this data is managed via CRUD operations on the REST-like API.
In total, the databases contain over five hundred items from four logic domains. 
Moreover, besides the predefined, ready to be used items, and perhaps more importantly, 
the backend application layer also serves as a tool for students or instructors to add items of their own interest 
to the database and use the reasoning and visualization tools in an approach to learning or teaching that fits 
their needs. The right panel in Figure~\ref{figure:main+editor} shows the formula editor GUI, 
details on how the GUI maps to CRUD actions are in Section~\ref{sec:managing-repositories}.
The server contains mainly databases of formulae libraries and models corpora in the four logic domains mentioned
and a repository of interactive reasoning tasks (dialogic/semantic games, etc.) using both formulae and models.

\textit{The Frontend User Interface}
The client side layer of the program uses HTML5 and CSS3 for structure respectively presentation 
and JavaScript for handling application logic as well as for implementing 
logic specific functionality.
The framework used to build the frontend application layer is Backbone 
enhanced with Marionette 
modules, together with their standard dependencies JQuery 
and Underscore~\cite{JQuery-Underscore-Backbone-Marionette}. 
For the graphical user interface we use Bootstrap 
and for rendering formulae in a browser MathJax. 
For displaying logical structure we use SVG 
and for interacting with reasoning tasks visualizations we use D3~\cite{Bootstrap-MathJax-SVG-D3}, 
particularly tree and force graph layouts.
Figure~\ref{figure:main+editor} presents the main user interface components
and Section~\ref{sec:managing-repositories} describes standard use scenarios for the GUI.
Figure~\ref{figure:syntax+tableau+models} illustrates some visualizations of logical structure
and Section~\ref{sec:interactive-visualizations} describes how these can be interactively 
generated, transformed and explored. 
The logic specific application component consists of several modules containing functionality 
handling formal aspects concerning the syntax and semantics of the logics 
and  dealing with their associated reasoning tasks.
These include a term unification package 
and an expression parser~\cite{PEG-unify}, 
a module for translating syntax into 
D3 tree layout, a tableau based theorem prover~\cite{ijcar} with a corresponding  module to translate the 
analytic tableau proof trace into a D3 force graph layout with a hierarchical tree structure,
and a (counter)model generator~\cite{phd} with a corresponding module for translating the (closed) tableau branches 
into (counter)example relational structures and further into a D3 force graph layout.

Further information about implementation details and the technology stack used are included in the electronic version of 
this paper and in the documentation available on~\cite{raeson}. 
The program was developed and tested using primarily Chromium in an Ubuntu desktop environment
and will run optimally in WebKit-based browsers and using a desktop format factor.
A live demo of the program is openly available online at~\cite{raeson}. 

\section{Managing Repositories of Formal Knowledge}
\label{sec:managing-repositories}

A first feature of the program consists in providing an interface for accessing and 
working with the following logic-specific repositories of formal knowledge: 
\begin{enumerate}
 \item {\it Formulae Libraries} provide an interface for accessing databases containing syntactic items from  
 various logical languages. Currently, these include:  
 Propositional Logic, Modal Logic, Predicate Logic, Quantified Modal Logic.
 \item {\it Models Corpora} provide an interface for accessing databases containing semantic items associated with
 various logics. These are, depending on domain: 
 truth assignments, relational structures, first-order models, QML models.  
 \item {\it Games Repositories} provide an interface for accessing databases containing interactive reasoning tasks.
 These are not entries for separate entities but merely documents associating, formulae and/or models with logic modules.
\end{enumerate}  
To facilitate access to repository items and reasoning tasks, the main interface layout has three component regions 
containing distinct information panels. 

\begin{figure}[t] 
\centering
\includegraphics[width=397px,height=190px]{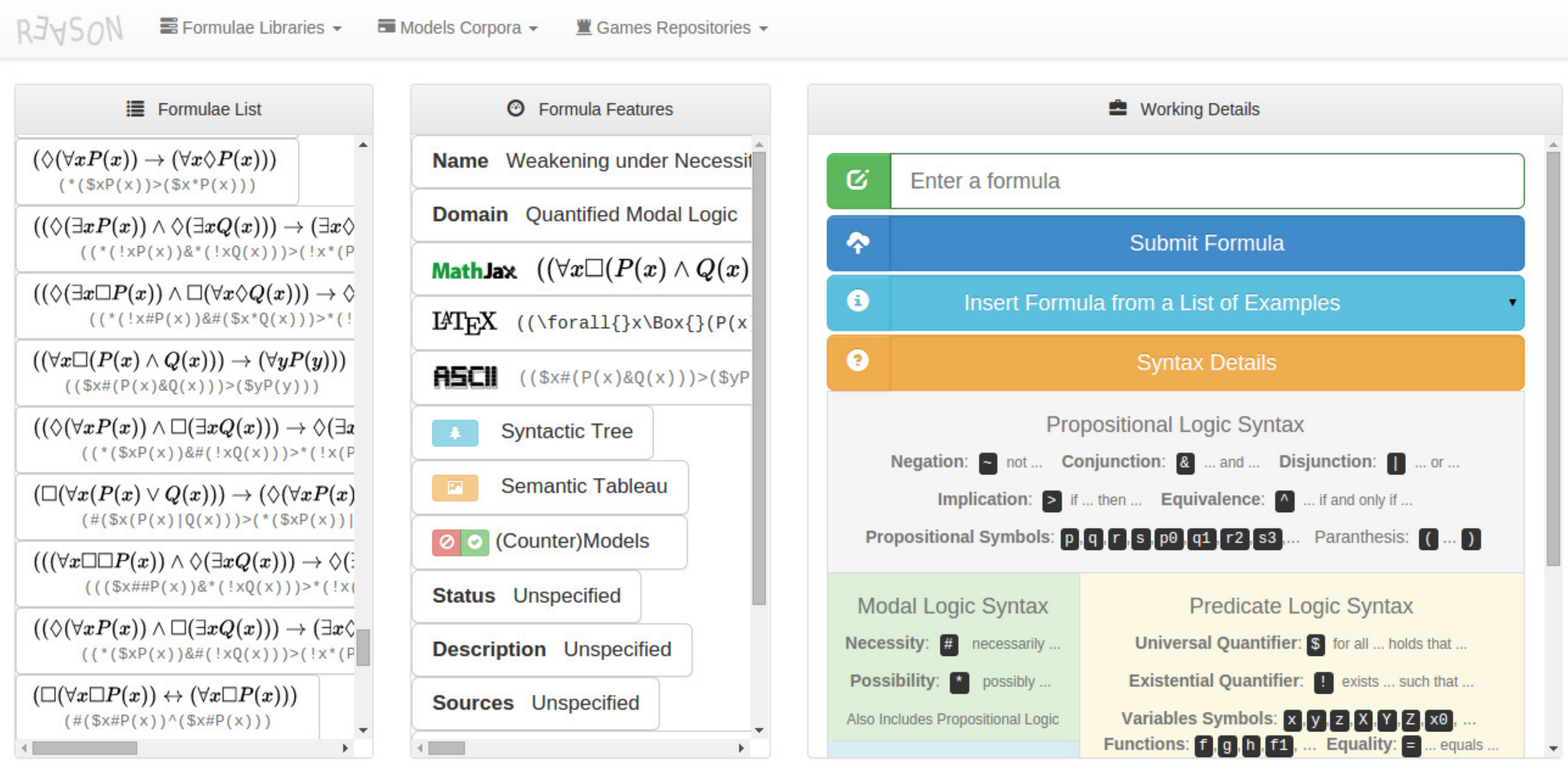}
\caption{Graphical user interface components: library view, formula features and formula editor {\sf } \label{figure:main+editor}} 
\end{figure}

The first panel, in the left in Figure~\ref{figure:main+editor}, is a collection view, 
its role is to visualize the repository content in an items list format. 
These can be either a formulae library or a corpus of models from those enumerated above.
The second panel, shown in the middle of Figure~\ref{figure:main+editor}, is an item view, 
its role is to display item features. These features are specific to each repository type and logic domain and link 
to various reasoning tasks, additional natural language information or further symbolic representations for a particular 
repository item selected from the items list panel.
The third panel, in the right of Figure~\ref{figure:main+editor}, is also an item view. 
It will display, depending on the chosen feature, the specific visualization content and/or an interface for interaction.  
This can be, fore example, a SVG canvas, a editor form, a table, etc.
The specific details of each interactive reasoning task are described in Section~\ref{sec:interactive-visualizations}.
The subviews are interlinked and in a usual workflow will be navigated from left to right by first choosing an 
item of interest from a collection then a specific feature and finally working with its interactive visualization 
content in the working details panel. The main navigation menu also provides links to instances of the three layout regions.
More detailed descriptions of possible workflows applied to concrete examples are presented in Section~\ref{sec:dydactic-touch}.

\textbf{Working with the Formula and Model Editors }
The first level of feedback guided interaction with logical structure takes place starting from syntactic aspects 
and is mediated by the formula and model editors. The right side panel in Figure~\ref{figure:main+editor} illustrates
the editors' user interface. 
This includes examples of well formed entities for each logic domain
and detailed information about the formation rules. 
Feedback is provided as the user types 
and parser errors are displayed whenever an incorrect string is entered.
A modular parsing expression grammar is used for input validation. This extends 
naturally from logical fragments to more expressive logics, from object language to the 
specification metalanguage, and from formulae to model syntaxes. 
Section~\ref{sec:code} contains an illustration of the the grammar modules. 
A concrete editor use case example is given below in Section~\ref{sec:dydactic-touch}.
Further details are included in the electronic version of 
this paper and in the program documentation available on~\cite{raeson}.

\textbf{Working with Repository Item Features}
A second level of feedback guided interaction with logical structure is mediated by reasoning tasks visualizations, 
accessed through specific item features. The main feature categories are:
\begin{enumerate}
 \item {\it Reasoning Tasks} is the main category, it contains various entries such as: visualizing syntactic construction trees, 
 building semantic tableaus, generating countermodels and models, etc.  
 Each task is described in detail in Section~\ref{sec:interactive-visualizations}. 
 \item {\it Item Information} contains various categories of information about the chosen items in natural language, such
 as name, description, sources, domain, etc. 
 \item {\it Symbolic Representation Formats} used for reference, linking, translation, etc.
\end{enumerate}  
Clicking on a formula feature will display its corresponding working details in the right hand side panel
of the main layout presented in~Figure~\ref{figure:main+editor}. 
While the first two panels are used for organization and navigation of formal repositories, 
the working details panel is used to host and manage most of the interactive aspects by   
coupling interface actions with a feedback loop allowing the user to generate and/or explore the formal structure 
underlying and propelling various logical reasoning tasks.

\section{Interactively Visualizing Reasoning Tasks}
\label{sec:interactive-visualizations}

The main interactive feature of the program is visualization of logical structure. 
This takes place for several kinds of analysis and reasoning tasks specific to each logic domain.
In this section we describe how to use the main reasoning tasks.

\begin{figure}[t] 
 \centering
\includegraphics[width=397px,height=135px]{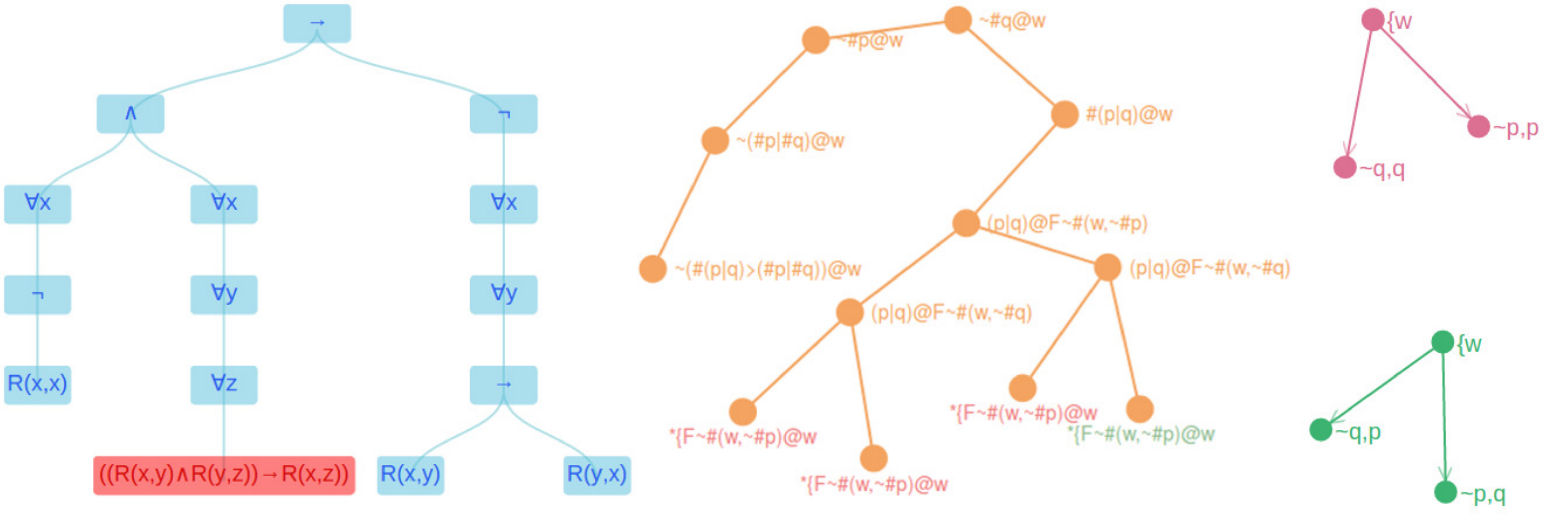}
\caption{Examples of interactively visualizing logical structure in reasoning tasks: syntactic construction tree (left), 
semantic tableau proof trace (middle),
generated counterexamples/models~(right) \label{figure:syntax+tableau+models}} 
\end{figure}

\textbf{Syntax Construction Trees}
A first analysis task is the interactive generation of syntactic construction trees.
This uses the syntactic formation rules of each logic domain to generate a tree layout of the formula and
all its subformulae, all the way down to its atomic components.
The left side of Figure~\ref{figure:syntax+tableau+models} illustrates a syntactic tree layout generated in this 
fashion for a first order logic formula. The interface distinguishes between nodes containing fully extended components 
and the nodes containing further analyzable components. 
The fully extended tree branches and the nodes containing atomic syntactic elements are marked with a blue color
while the not yet fully analyzed nodes are colored in red, the user can further extend these nodes by clicking on their content. 
In the illustration the second branch is not yet fully decomposed, this is marked in red 
and can be interactively extended further by clicking on the terminal subformula. 
Concrete teaching context are included in Sections~\ref{sec:dydactic-touch},~\ref{sec:classroom}.

\textbf{Semantic Tableaus}
A second analysis task is the interactive generation of semantic tableau proofs. 
This uses the semantic decomposition rules specific to each logic domain to generate a force tree layout
by analyzing the head formula in a branch and so on in all resulting tableau branches, all the way down to 
exhausted branches. 
The middle region of Figure~\ref{figure:syntax+tableau+models} illustrates a semantic tableau generated in this way
for an example formula from the modal logic domain.
The interface distinguishes between nodes representing fully expanded branches and the nodes standing for 
branches that can be further expanded i.e. contain formulae that can be further analyzed. The layout also 
distinguishes between closed, i.e. containing contradictory formulae, and open branches, i.e. fully expanded but not
containing a formula and its negation.
Nodes are labeled by the head position formula in the branch, this is the formula logically analyzed and decomposed 
in that expansion step. The user can interactively display the full content of a tableau branch by hovering over 
any node of the tableau. Branches which are not fully expanded yet can be extended by clicking on a terminal node.
This triggers the process of logical decomposition on the head formula in the branch and adds the resulting branch(es) 
to the tableau.
Clicking on an expanded node reverts the tableau expansion step and removes the outgoing branch(es).
For modal logics in addition to decomposition rules for formulas in the logical object language 
the tableau construction process also uses a specification metalanguage and corresponding decomposition rules
for hybrid logic with nominals. 
These decomposition rules are based on applying metanominals to possible worlds as Skolem terms constructed from 
formulae containing modal operators and using relational properties of accessible 
successors during tableau expansion. The theoretical background behind this technique has been described 
in full detail in~\cite{ijcar}, and the logical grammar used for parsing hybrid logic formulae in the 
specification metalanguage is also included for reference in 
Section~\ref{sec:code}. Concrete teaching context are included in Sections~\ref{sec:dydactic-touch},~\ref{sec:classroom}.

\textbf{Countermodel Generator}
A third reasoning task consists in generating countermodels for a given formula. 
This starts from the proof trace generated by a semantic tableau and from each open branch in the terminal nodes
of the tree layout extracts a corresponding model. Hybrid logic formulae with nominals are used for expressing 
relational components of a model inside the specification metalanguage. These are then translated in a force graph layout 
representing a relational structure which is displayed for visualization and can be further interactively inspected. 
The right hand side of Figure~\ref{figure:syntax+tableau+models} presents two graphs generated in this fashion.
The interface distinguishes counterexamples falsifying the formula, which are colored in red, from models satisfying the 
formula, which are colored in green. The satisfying models are obtained in a similar way starting from a complementary formula.
Each node of the graph is labeled with a relevant extensional component, for modal logic this is the world's propositional
valuation. The user ca obtain further information about the graph by hovering over the nodes. This will reveal the metanominal
corresponding to each possible world. This nominal is a Skolem term generated during tableau construction and contains further
data about the stage in the proof trace it was introduced and the formula and other metaterms it was generated from.
Concrete teaching contexts are included in Sections~\ref{sec:dydactic-touch},~\ref{sec:classroom}. 

Several additional reasoning tasks are described and/or exemplified 
in Sections~\ref{sec:dydactic-touch} and~\ref{sec:classroom}, 
in the extended version of this paper and the documentation available on~\cite{raeson}.
These reasoning tasks are either specific to a particular logic domain, 
such as Truth Tables for Propositional Logic, or only relevant for corpora of models, 
such as Model Inspection etc.

\section{The Didactic Touch} \label{sec:dydactic-touch}
In this section we will present a series of concrete use case scenarios: 
in order to illustrate the way in which the program can be useful for didactic purposes.

\textbf{Use Case Scenario 1: Exploring Formal Knowledge Repositories} This use 
case will step through a typical repository navigation.
Step~(1): From the {\sf Main Navigation Menu} select {\sf Formulae Libraries $>$ Modal Logic}, alternatively, you can click on the 
{\sf Modal} link from the {\sf Go to a Formula Library} green div in the {\sf Items List} panel in the left of the main layout. 
The left side panel will display the content of the {\sf Modal Logic Library}. 
Step~(2): Search in the list the formula named {\sf Distribution Axiom (K)}, alternatively, you can hit 
{\sf ctrl+f} and type the formula's ascii content {\tt (\#(p>q)>(\#p>\#q))} or its name,
as a shortcut to finding the formula faster (using native browser search). 
When you click on the formula, the middle panel will display the formula's features.
Step~(3): Inspect the list of available features for the formula listed in the middle panel and
find either the {\sf Name} or the {\sf Domain} item. When you click on the chosen feature the right hand side panel 
will display additional information about the item of interest. In this case this will be additional 
information in natural language about the formula. 
Item features related with reasoning tasks are explored in the next use case.

\textbf{Use Case Scenario 2: Interactively Visualizing Reasoning Tasks}
This use case will present step by step a possible scenario of exploring reasoning tasks. 
Repeat steps (1) to (3) from the previous use case, but this time use {\tt (\#(p|q)>(\#p|\#q))} to find formula 
named {\sf Distribution Axiom (F)}, note the (F) at the end which distinguishes it from the previous example. 
Step~(4): Click on the {\sf Syntactic Tree} feature and explore the construction tree by clicking on the nodes of the tree.
Note that the syntactic construction of the formula is the same with the one displayed in the previous example 
(both formulae generate the same tree structure). 
Step~(5): Click on the {\sf Semantic Tableau} feature and fully expand all the tableau branches by clicking on the nodes 
of the tree layout. By hovering over the nodes in the tableau you can see the entire structure of the branch 
containing all the extended formulae up to that point. Continue until all the formulae are logically decomposed
and every branch of the tableau is exhausted. Note that the semantic tableau constructed in this case is 
different from the one for the previous formula. Find the green colored open branch that makes the difference and 
inspect its content, try to imagine a counterexample starting from it. 
Step~(6): Click on the {\sf Model Generator} feature and scroll the panel of models generated in the {\sf Working Details}
panel until you find one that is colored in red. The nodes in the graph represent possible worlds and are labeled 
by their propositional valuation. Hovering over possible worlds in the models will reveal their 
construction nominal, compare it with the open branch found in Step~(5).  

\textbf{Use Case Scenario 3: Update Formal Knowledge Repositories} 
In the~{previous} use case scenario you have found a counterexample to a formula labeled as an axiom, 
now you are going to fix this. If needed, go through steps~(1-6) in the previous use case to obtain
the open tableau and the countermodel illustrated in Figure~\ref{figure:syntax+tableau+models} (middle, respectively, right).
Step~(7): Click on the {\sf Copy to Sandbox Library} feature. The {\sf Sandbox Library} repository will be displayed in 
the {\sf Items List} panel. Find the formula and click the {\sf Edit Formula Information} feature. The rightmost panel will
display a form with editable entries. 
Step~(8): Change formula's name from {\sf Distribution Axiom (F)} to 
{\sf Distribution Formula} (fails for disjunction). Optionally, add a description and sources, then
click the {\sf Submit Changes} button. 
Step~(9): Find the modified formula in the {\sf Sandbox Library}, click the {\sf Edit Ascii Content} feature.
The {\sf Formula Editor} will open with the formula's ASCII content available for editing.
Step~(10): Change all disjunctions in the formula to conjunctions, as indicated in the available instructions 
panels this is done by replacing all occurrences of~{\tt |}~with~{\tt \&}. Feedback will be provided as you type,
when the new formula is parsed correctly, click on the {\sf Submit Formula} button.
Eventually, repeat Step~(8): renaming the formula {\sf Distribution over Conjunction} and adding the description 
``This fails for disjunction''.

\textbf{Use Case Scenario 4: Semantic Tableaus Expansion Heuristics} 
Step~(1): Use the semantic tableau feature to find a counterexample model falsifying the modal formula 
$(\Box{}(\lnot{}p\lor{}\Box{}q)\leftrightarrow{}(\Box{}\lnot{}p\lor{}\Box{}q))$, {\sf Theorem (S5) 3} in the {\sf Modal Library},
using no more than seven expansion steps.
Step~(2): Can you do this with less than seven clicks? If yes, how many?
If no, what syntactic property of the formula can be used to prove this for the general case.

Several other use cases are possible including features such as: 
interactive visualization and inspection of models, linking formulae with structures through model checking tasks, 
playing dialogic or semantic games, etc. More use cases are described and explored 
in the electronic version of this paper and in the program documentation available on~\cite{raeson}.

\section{Classroom Settings Using the Tool}\label{sec:classroom}

In this section we describe several possible scenarios of using the program in a classroom setting 
and the didactic experience that has been accumulated.
The program can be used both during live classroom activities and also
remotely by students. The live teaching classroom settings can integrate the tool 
alongside traditional teaching methods also during theory exposition but mostly for solving exercises.
The following list describes typical exercise templates (concrete examples can be randomly selected from repositories):
\begin{description}
 \item[Exercise Model One] Given: A list of several formulae in symbolic notation (MathJax-\LaTeX) and a matching 
 list of formulae represented as syntax construction trees (D3-SVG). 
 Required: Establish the correct correspondence between the items in the two given lists.
 \item[Exercise Model Two] Given: A list of structures in extensional notation (MathJax-\LaTeX) and a matching 
 list of models represented as labeled graphs (D3-SVG force layout). 
 Required: Establish the correct correspondence between the items in the two given lists.
 \item[Exercise Model Three] Given: Syntax construction trees of several formulae with matching structure graphs 
 (pointed in the case of modal logics, i.e. with designated actual worlds). 
 Required: Decide for each pair if the (pointed) model satisfies or not the given formula.
 \item[Exercise Model Four] Given: A list of formulae in symbolic notation (MathJax) and a matching list of 
 (pointed) models in extensional representation (not as D3-SVG graphs). Required: For each formula in the list select 
 the (points)/structures that satisfy/falsify it.
 \item[Exercise Model Five] Given: A list of (pointed) models represented in graphical format (D3).  
 Required: For each structure in the list upload into the formulae library M formulae that are true/false 
 in the model (at the actual world) and having (modal) depth at least N. 
 \item[Exercise Model Six] Given: A list of (pointed) models represented in extensional format.  
 Required: For each structure in the list upload into the formulae library N formulae that are false/true 
 in the model (at the actual world) and having (modal) depth at least M. 
 \item[Exercise Model Seven] Given: A list of formulae represented as syntactic trees (D3-SVG). 
 Required: For each formula either upload into the corresponding model corpus a counterexample structure with at 
 least N possible worlds(/objects) or claim validity otherwise. 
 \item[Exercise Model Eight] Given: A list of formulae represented symbolically (MathJax-\LaTeX).
 Required: For each formula either find and write down in extensional form a counterexample structure with at 
 least M possible worlds(/objects) or claim validity otherwise.
\end{description} 
The previous exercise models can be used in classroom activities, progress evaluations and final exams. 
To asses the didactic relevance of the program a nonequivalent groups design can be used 
to measure the average grades in student groups that used the program or not. 
Another measure can be the response time for accurate answers.
Early adopters have (only recently) started to use the program in teaching activities%
\footnote{
Dr.~Adrian Ludu\c san is using {\sf R\rotatebox[origin=c]{180}{E}\rotatebox[origin=c]{180}{A}SON} in teaching activities for logic courses 
at ``Babe\c s-Bolyai'' University} 
 and relevant data can be obtained by logging user interaction based on the previous list of exercise templates. 
As the design of the exercises suggests, the main testing
parameter is the usefulness of interactively visualizing logical structure in reasoning tasks.
Preliminary results \cite{ludusan} have been used so far only for internal feedback loop. These show that the test group significantly 
outperforms the control group for exercise model one, which is as expected. For exercise model three the grades average 
is also better in the test group, which is also not surprising. The really interesting result is that the 
test group outperforms the control group also for exercise model four (no visualizations).
Moreover, when comparing averages for exercise model three with those for exercise models four, 
inside the same group, the former significantly outperform the later. This is even more relevant as 
this outcome holds not only for the test group, but also for the control group. All these suggest 
that interactive visualizations of logical structure in reasoning tasks can be didactically relevant.
At the moment, not enough data is available for the remaining exercise models
in order to warrant conclusion validity as a rigorous statistical experiment 
(more results will become available online as more data is collected).

A final didactic experience that deserves to be mentioned is that the tool facilitates types of exercise that might 
be impractical (if not impossible) to stage using traditional teaching media (like pen and paper or even chalk and a wide blackboard).
Consider an example:
\begin{description}
 \item[Example 1] Given: {\sf Rule of consensus, exception\,1} from the propositional logic formulae library.  
 Required: Find a counterexample structure using the semantic tableau method.
\end{description}

\section{Conclusions}
\label{sec:conclusion}
The paper presented a software tool for analysis and interactive engagement 
in various logical reasoning tasks.
The main feature of the program is to provide the means to intuitively visualize and 
interactively generate the logical structure that propels customary 
logical reasoning tasks. 
A second feature consists in providing an interface for working with 
logic-specific repositories of formal knowledge. 
Starting from this we showed how both aspects have didactic potential and can be integrated in 
teaching activities to provide an engaging learning experience. 
Overall, this creates an ecosystem of interactive software tools modeling reasoning 
tasks driven by automated visualization of logical structure emerging in logics that are often 
underrepresented in or overlooked by existing approaches. 
We also illustrated how this approach can be useful for didactic purposes.
Besides presenting a possibly new logical reasoning software suite, the paper also revisited classical challenges 
for teaching logic and reasoning skills using a completely open source software stack for 
development, testing, deployment and hosting. 
The resulting program is openly available online \cite{raeson}, 
it is implemented completely platform independent and can be run in any modern browser.

Topics of further work include adding relational theories to the modal metatheory and customizing 
the modal tableau prover. 
Extending the modeled reasoning tasks
to currently uncovered logic domains. Adding additional layers of interactivity to the interface 
and a customizable strategy for tableau expansion. Adding dynamic logic elements, etc. 
\section{Logical Syntax Grammars}\label{sec:code}
\begin{lstlisting}[caption={Atomic Grammar Components (common for both Formulae and Model syntaxes)},
                   numbers=left,stepnumber=1,basicstyle=\tt\scriptsize,label=listing:3,
                   captionpos=t,abovecaptionskip=-\medskipamount]
propsym "proposition" = chars:([p-s0-9_\-]+) {return chars.join("")}
nominal "nominal" = chars:([i-k0-9_\-]+) {return chars.join("")}
metanominal "metanominal" = chars:([lt-w0-9_\-]+) {return chars.join("")}
predicate "predicate" = chars:([P-SA-EMN0-9_\-]+) {return chars.join("")}
function "function" = chars:([f-h0-9_\-]+) {return chars.join("")}
variable "variable" = chars:([X-Zx-z0-9_\-]+) {return chars.join("")}
constant "constant" = chars:([a-emn0-9_\-]+) {return chars.join("")} 
\end{lstlisting}
\begin{lstlisting}[caption={Formula Syntax: Complex Grammar Components},
                   numbers=left,stepnumber=1,basicstyle=\tt\scriptsize,label=listing:1,
                   captionpos=t,abovecaptionskip=-\medskipamount]
start = metaformula / formula
formula "formula" = atomic / unary / binary / quanty / metaterm2form
metaformula "metaformula" = formula "@" metaterm  
quanty "quantified" = "(" quantifier variable formula ")"
quantifier "quantifier" = "$" / "!" 
binary "binary" = "(" formula "&" formula ")" / "(" formula "|" formula ")"    
                / "(" formula ">" formula ")" / "(" formula "^" formula ")"
unary "unary" = "~" formula / "#" formula / "*" formula  
atomic "atomic" = nominal / propsym / equality / predicate "(" term ")" 
                / predicate "(" term "," term ")"
equality "equality" = "(" term "=" term ")"
term "term" = constant / variable / function "(" term ")" 
            / function "(" term "," term ")"
metaterm "metaterm" = metanominal / metafunction "(" metaterm "," formula ")"
metaterm2form "metaterm2form" = "{" metaterm // reductionaxiom = "<" atomic
metafunction "metafunction" = "F*" / "F~#"          
\end{lstlisting} 
\begin{lstlisting}[caption={Model Syntax: Complex Grammar Components},
                   numbers=left,stepnumber=1,basicstyle=\tt\scriptsize,label=listing:2,
                   captionpos=t,abovecaptionskip=-\medskipamount]
start = model
model "model" = left:world ";" right:model / world
world "world" = metanominal ":" extensionnotationslist
extensionnotationslist "extensionnotationslist"
  = left:extensionnotation right:extensionnotationslist / extensionnotation / ""
extensionnotation "extensionnotation" = valsymb valuation / relsymb relations
  / domainsymb domain / monadicsymb monadicext / diadicsymb diadicext
extsymb "extsymb" =  valsymb / domainsymb / monadicsymb / diadicsymb
valsymb "valsymb" = "VL"
valuation "valuation" = left:literal "," right:valuation / literal / ""
relsymb "relsymb" = "RL"
relations "relations" = left:metanominal "," right:relations / metanominal / ""
literal "literal" = atomic / "~" atomic:atomic {return "~" + atomic;}
atomic "atomic" = nominal / propsym / equality / predicate "(" groundterm ")" 
  / predicate "(" groundterm "," groundterm ")"
domainsymb "domainsymb" = "DO"
domain "domain" = left:groundterm "," right:domain / groundterm / ""
monadicsymb "monadicsymb" = "MP"
monadicext "monadicext" = predicate "{" objectslist "}"
monadicextslist "monadicextslist" = left:monadicext "," right:monadicextslist / monadicext / ""
objectslist "objectslist" = left: groundterm "," right: objectslist / groundterm / ""                                  
diadicsymb "diadicsymb" = "DP"
diadicext "diadicext" = predicate "{" pairslist "}"
diadicextslist "diadicextslist" = left:diadicext "," right:diadicextslist / diadicext / ""
pairslist "pairslist" = left: tuple2 "," right: pairslist / tuple2 / ""
tuple2 "tuple2" = "(" groundterm "," groundterm ")" 
groundterm "groundterm" = constant / function "(" groundterm ")"
                        / function "(" groundterm "," groundterm ")"
\end{lstlisting}

\end{document}